# Fermi-Bose mixtures and BCS-BEC crossover in high-T$_c$ superconductors

**Maxim Yu. Kagan** [1, 2*], **Antonio Bianconi** [3,4,5]

[1] National Research University "Higher School of Economics", ul. Myasnitskaya 20, 101000, Moscow, Russia, mkagan@hse.ru
[2] P.L. Kapitza Institute for Physical Problems, Russian Academy of Sciences, ul. Kosygina 2, 119334 Moscow, Russia, kagan@kapitza.ras.ru
[3] Rome International Center for Materials Science, Superstripes, via dei Sabelli 119A, 00185 Roma, Italy ;
[4] Institute of Crystallography, CNR, Via Salaria Km 29.300, Monterotondo, I-00015 Roma, Italyia
[5] National Research Nuclear University MEPhI (Moscow Engineering Physics Institute), 115409 Moscow, Russia

*Correspondence: kagan@kapitza.ras.ru, mkagan@hse.ru
antonio.bianconi@ricmass.eu

**Abstract:** In this review article we consider theoretically and give experimental support to the models of the Fermi-Bose mixtures and the BCS-BEC crossover compared with the strong-coupling approach, which can serve as the cornerstones on the way from high-temperature to room-temperature superconductivity in pressurized metallic hydrides. We discuss some key theoretical ideas and mechanisms proposed for unconventional superconductors (cuprates, pnictides, chalcogenides, bismuthates, diborides, heavy-fermions, organics, bilayer graphene, twisted graphene, oxide hetero-structures), superfluids and balanced or imbalanced ultracold Fermi gases in magnetic traps. We build a bridge between unconventional superconductors and recently discovered pressurized hydrides superconductors H3S and LaH10 with the critical temperature close to room temperature. We discuss systems with line of nodal Dirac points close to the Fermi surface and superconducting shape resonances, and hyperbolic superconducting networks which are very important for the development of novel topological superconductors, for the energetics, for the applications in nano-electronics and quantum computations.

### 1. Introduction

The seminal discovery in 1986 of high-$T_C$ superconductivity at 36 K in a doped cuprate perovskite $La_{2-x}Ba_xCuO_4$ by K. A. Muller and J. G. Bednorz in IBM Zurich [1,2] determined the shift of research for room-temperature superconductors from metallic alloys to complex perovskites. In the first 10 years the critical temperature of gradually increased from 36 K in lanthanum family of superconducting cuprate perovskites till 90 K in yttrium family $YBa_2Cu_3O_{7-\delta}$ and finally reaching 160 K in $HgBa_2Ca_2Cu_3O_8$ under a pressure of $P$ = 350 Kbar. This extraordinary progress promised a technological revolution in electronics, and energetics and provided the driving force for a tremendous progress in the development of many body theories for condensed matter physics, inorganic chemistry and material science. Today we see the growing interest on quantum complex materials. The conventional superconductivity in low-temperature superconductors is described by the BCS weak coupling theory in a homogeneous metal with pairing on a large Fermi surface with high Fermi energy. The unconventional high temperature superconductors are characterized by the breakdown of the standard BCS approximations: 1) a single electronic component, 2) high electron density and high Fermi energy $E_F$; 3) low values of the ratio $\omega_0/E_F$ between the energy cut off of the attractive interaction $\omega_0$ and the Fermi energy 4) large Fermi surface and high Fermi momentum $k_F$; 5) superconducting energy gap much smaller than the Fermi energy $\Delta \ll E_F$; 6) large ratio between the coherence length $\xi_0$ and the average distance between electrons.

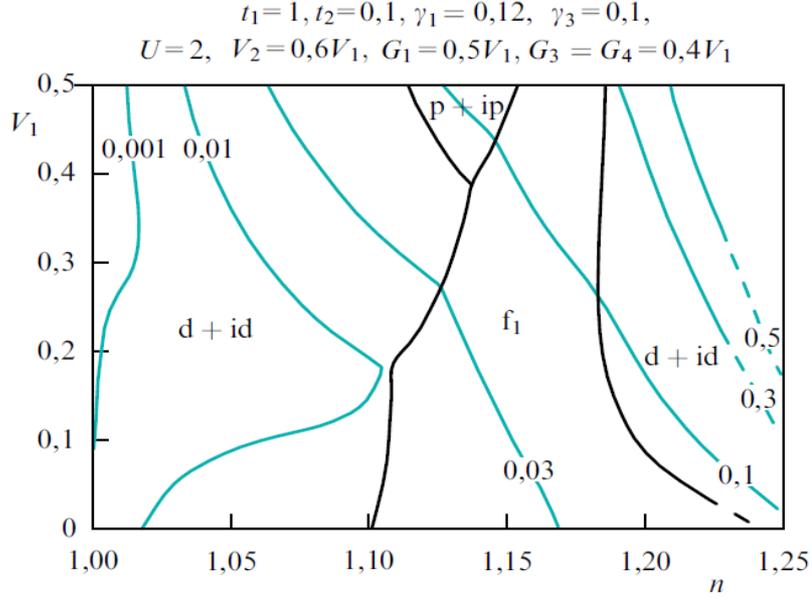

**Figure 1.** Phase diagram of superconducting state in the AB structure of the bilayer graphene [20] shown as a function of the variables $n - V_1$, where $1 < n < 1.25$ is the carriers density, $t_1, t_2$ and $\gamma_1, \gamma_2$ are intralayer and interlayer hoppings, $U$ is onsite Hubbard repulsion, $V_1, V_2$ and $G_1, G_3, G_4$ are intralayer and interlayer Coulomb repulsions. The leading SC instabilities correspond to chiral $p + ip, d + id$ and $f_1$-pairing. For all the points on the same thin line the value of the coupling constant $|\lambda| = 1/\ln(E_F/T_C)$ is constant.

Soon after the discovery of Müller, Anderson and Schrieffer advanced different unconventional mechanisms for high temperature superconductivity (see e. g. [3, 4, 5]). Many fruitful ideas were introduced in the first years of HTSC by the leading theorists in USA, Europe, Japan and worldwide (see e. g. [6,7,8,9,10,11,12,13,14,15,16]). One of the authors of the present review was very active in unconventional mechanisms focusing on the low density electron systems [17,18,19,20,21,22,23,24] based on the generalization of Kohn–Luttinger ideas [25,26] in purely repulsive Fermi systems. In the framework of the Fermi gas model with hard-core repulsion [27], repulsive-U Hubbard model [28] and the Shubin–Vonsovsky model [29, 30] on different type of 3D and 2D lattices during last 30 years of intensive research we predicted p-, d-, f- and anomalous s-wave pairing in various materials such as the idealized and bilayer graphene (see Fig. 1), superconducting pnictides and organic superconductors and superfluid $^3$He.

It has been shown that the critical temperature of the superconducting transition can be substantially increased at low densities by considering spin-polarization or the two-band superconductivity scenario [31,32,33]. These results proved to be very important for related systems as in polarized $^3$He - $^4$He mixtures, imbalanced Fermi-gas in magnetic trap as well as for heavy fermion superconductors and other mixed valence systems described by the two-band Hubbard model with one narrow band [34,35,36,37]. Together with T.M. Rice we also considered a superconductivity scenario in 2D t-J model at low and intermediate electron density [38, 39] with van der Waals interaction which corresponds to strong onsite Hubbard repulsion and weak intersite antiferromagnetic (AFM) attraction. The phase-diagram of this model presented in Fig. 2 shows regions of superconducting phases with different pairing symmetry: p-wave, s-wave and $d_{x^2-y^2}$-pairing. For small values of $J/t$ the phase-diagram of the 2D t-J model becomes equivalent to the phase-diagram of the 2D Hubbard model and according to the results [17-19] is unstable at low electron densities towards p-wave superconductivity of the Kohn-Luttinger [25] type. At large values of $J/t \geq 2$ (which usually are not realized in cuprates) the system becomes unstable towards extended s-wave pairing. For the set electron densities (optimal doping) and the values of $J/t \sim 1/2$ we get the critical temperature of the d-wave pairing typical for high-$T_C$

$$T_C \sim E_F \exp(-\frac{\pi t}{2 J n_e^2}) \sim 100 K . \quad (Eq. 1).$$

This result was generalized by N.M. Plakida et al [40,41] on the opposite case of small hole densities $x = (1 - n_{el}) \ll 1$ using the diagrammatic technique of the Hubbard operators for the pairing of two spin polarons. The same order of magnitude $T_C$: 100 K for the critical temperature can be obtained for the d-wave pairing in the Shubin-Vonsovsky model in the intermediate coupling case for the set of parameters $U \approx 3t, n_{el} \approx 0.9$ [20,21,24] on the 2D square lattice.

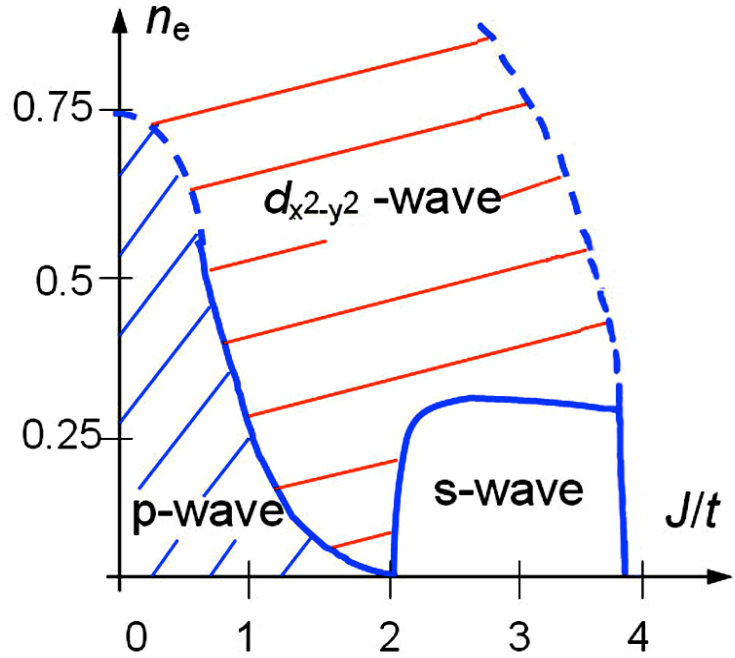

**Figure 2.** The phase-diagram of superconducting state of 2D t-J model [38,39].

We would like to summarize here both the early stage [42-53] and recent [54-58] theoretical results of the Russian research school in high-$T_C$ investigations

The proposed mechanisms of superconductivity in high-$T_C$ cuprates range from BCS scenarios involving electron-phonon or electron-electron interactions, with extended pairs [59] or BEC scenarios with local pairs [60, 61], with condensates of s-wave or d-wave [4,6,12,8] in models of the normal phase described by Landau Fermi liquid [62, 63], Luttinger [64] or marginal [13] Fermi liquid (correspondingly with spinons and holons instead of fermionic quasiparticles), spin-charge separation [3,6,12,64,65] or spin-charge confinement [11,65], with the presence of pseudogap [66] in underdoped state, including interplay between superconductivity and antiferromagnetism [67], formation of stripes [14], the role of the second layer and the c-axis plasmon mode [15] have been subject of an intensive debate.

In 2015 following theoretical predictions [68, 69] for high Tc in the high pressure metallic phase $H_3S$ the record [70] for the superconducting temperature of 203 K was reached. The authors of the experimental breakthrough claimed that the discovered superconductivity could be explained by e conventional strong-coupling version of the BCS theory [71] constructed by Migdal and Eliashberg [72, 73] used to predict the high Tc phase in the dirty limit [69, 70]. It was rapidly shown that by changing the pressure to reach the maximum critical temperature the chemical potential in H3S is driven to an electronic topological transition known as a Lifshitz transition for the appearing of a new piece of a Fermi surface [74,75]. In this regime the Migdal approximation breakdown [76] and it is not possible to apply the BCS theory to explain the emergence of high temperature superconductivity. It has been shown that the emergence of high temperature superconductivity in a multiband scenario near a Lifshitz transition has been described by numerical solution of Bogoliubov equations [77] and can be qualitatively predicted in the limit of a steep –flat scenario [79] where the energy dispersion of the appearing band is pushed down to zero as for the case of an infinite effective mass for strong correlated localized states [78]. An essential point of this proposed scenario for the emergence of room temperature superconductivity is the theoretical demonstration that the Lifshitz transition for the appearing of a strongly correlated band is not of 2,5 order but becomes of first order with appearing of a frustrated phase separation [79], interface superconducting [80] and a hyperbolic space of filamentary pathways in two dimensional systems like it was observed in cuprate perovskite superconductors at optimal doping [81].

Recent papers on pressurized hydrides under high pressure support early papers [82] and support the multigap superconducting scenario in the proximity of a Lifshitz transition [83].

Quite recently two experimental groups [84,85] reported a discovery of even higher critical temperatures $T_C$: (250–260) K in lanthanum superconducting hydrides $LaH_{10\pm x}$ at high pressures *P:* (170–190) GPa. In this compound according to the Density Functional Theory (DFT) the host atom of La is at the center of the cage formed by hydrogen atoms. Moreover the authors of [86] on the basis of numerical calculations predict the topological Dirac nodal line in $LaH_{10}$ near the Fermi energy $E_F$. Note that room critical temperatures (RTSC) of the order of 300 K or higher in metallic hydrogen for very high pressures *P:* (170–190) Mbar and in hydrogen dominant metallic alloys (at smaller pressures) were predicted by Ashcroft [87, 88]. The large values of $T_C$ in Ashcroft estimates were facilitated by the high phonon frequencies governed by light mass of hydrogen ions.

Note also that profound numerical calculations of the optimal crystalline structure performed by Brovman, Yu. Kagan and Kholas [89, 90] show the regions of 3D isotropic phase, quasi-2D planar phase and quasi-1D filamentary phase on the *P–T* phase-diagram of metallic hydrogen. In particular they predict at relatively low pressures and *T*=0 a strongly anisotropic phase with the proton filaments embedded in electron liquid. The filaments can move almost freely relative to each other in the longitudinal direction similar to vortices in superfluid ⁴He [91] (where we have a vortex crystal in perpendicular direction to the vortex lines and a free superfluid in parallel direction to the vortex lines). In the same time the filaments form a rigid triangular lattice in the direction perpendicular to them which resembles Abrikosov vortex lattice in type-II superconductors [92]. Similarly in the planar phase the hydrogen layers can move almost freely relative to each other similarly to the smectic liquid crystals or graphite layers.

**2. Superconductivity in the Fermi-Bose mixture model. Theory.**

Let us stress that the light mass of hydrogen ions H⁺ not only guarantees the high values of Debye frequency but also leads to a large kinetic energy of zero-vibrations and thus to a large value of DeBoer parameter measuring the ratio of kinetic delocalization energy and potential energy and responsible for the quantumness of the system [93,39]. This fact brings a highly mobile ionic subsystem in metallic hydrogen (which resembles superionic crystal but at low temperatures in our case) close to the limit of the quantum crystal [93,39]. Note that a high mobility of filaments or layers can promote also rather large values of Lindemann parameter (Lindemann number [93,39]). It measures the ratio of the root mean square of the displacement of ion to the interionic distance in metallic hydrogen or in hydrogen dominant alloys and shows that our system is close to the quantum melting limit. In principle, according to the ideas of Andreev, Lifshitz the ionic lattice of the quantum crystal can become superfluid at low temperatures (due for example to a flow of vacancies or other defects relative to the lattice or due to delocalization of fillaments themselves which can form the macroscopic wave function). This phenomenon is nowadays called a supersolidity of the quantum crystal [94]. Experimental indications on available today scenarios for quantum supersolid are discussed by Bianconi group in [81]. The experimental support for filamentary supersolids can be found in [76, 78] having in mind not only "new" hydrogen dominant superconductors but also the "old" cuprate perovskites.

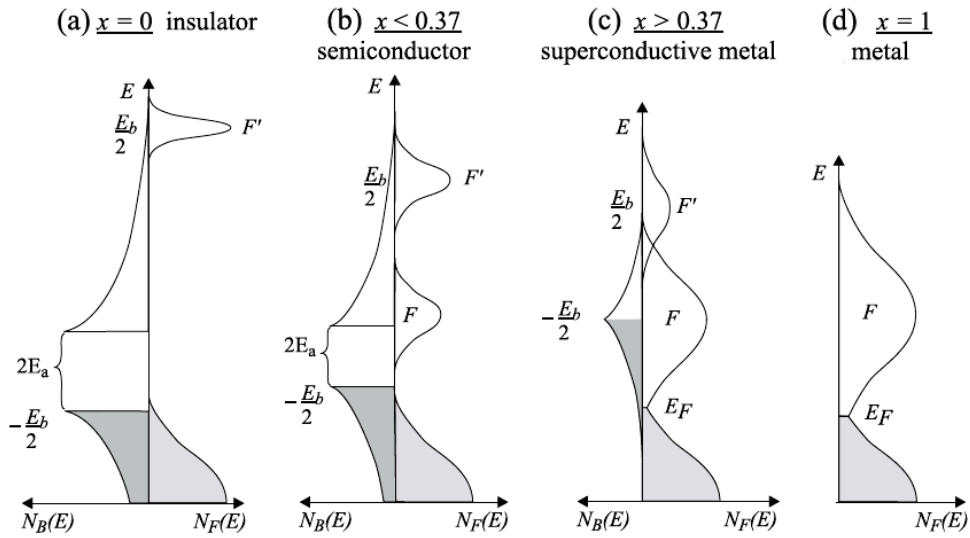

**Figure 3.** Interplay between Fermi and Bose subsystems in superconducting $Ba_{1-x}K_xBiO_3$ [111, 112, 21]: (**a**) insulator, $x = 0$; (**b**) superconductor, $x < 0.37$; (**c**) superconductive metal, $x > 0.37$; (**d**) metal, $x = 1$.

In analogy with neutron stars we can describe the ionic subsystem on the language of biproton superfluidity which coexists with the BCS Eliashberg superconductivity of the electron subsystem (probably modified on frequency dependence of effective mass and nonconstant density of states [95]).

These qualitative considerations press us to consider the superconductivity in metallic hydrogen on the level of Fermi-Fermi mixture [96] of protons and electrons (as in plasma physics or in the two components model with one very narrow band [36, 37]) or on the level of Fermi-Bose mixture of Cooper pairs and protons or even of Bose-Bose mixture [97] of Cooper pairs and biprotons. Note that the ideas of Fermi-Bose mixture were very fruitful in low temperature physics describing the search for fermionic (BCS) superfluidity in dilute 3D and 2D solutions of $^3$He in superfluid $^4$He [34,98,99,100] as well as $^6Li - ^7Li$ or $^{40}K - ^{87}Rb$ mixtures of ultracold atoms [101,102,103] in restricted geometry of magnetic and dipole traps or on the optical lattices.

Note that in ultracold Fermi gases there is a very effective way to change the magnitude and the sign of the interaction (more precisely of the s-wave scattering length) between the atoms. This can be done in the resonance magnetic field, in the regime of Feshbach resonance [104] which is important for the physics of BCS-BEC crossover in quantum gases. The Fermi-Bose mixture model corresponds to the two-channel description of the Feshbach resonance [105]. Here we have the attractive Majorana exchange term which transforms the (extended) Cooper pair consisting of two fermions in one channel into a real boson (a local pair or dimer) in the other channel.

In the physics of superconductors the model of Fermi-Bose mixture was firstly proposed by T. D. Lee et al. [106] and Micnas et al. [107]. for superconducting cuprates. In this model the Majorana exchange term which transforms the Cooper pair in a boson, is present. Later on Larkin, Geshkenbein and Ioffe [108] phenomenologically advanced this model on the level of Ginzburg–Landau theory [109] for 2D electron system described by Hubbard type models close to van Hove singularities and for two-leg ladder systems [110].

Note that the Fermi-Bose mixture model with an additional requirement of the spatial separation of Fermion and Boson component captures thermodynamic and transport properties as well as a lot of essential features on the phase diagram of superconducting bismuth oxides $Ba_{1-x}K_xBiO_3$ [111,112]. The local pairs are formed in these materials in $BiO_6$ clusters (see Fig. 1). The critical temperature $T_C$ 36 K corresponds here to the coherent tunneling of local pairs between neighboring Bose clusters through the effective barriers formed by Fermi clusters.

The generic Hamiltonian of the Fermi-Bose mixture typically has the form

$$H = H_F + H_B + H_{BF} \quad \text{(Eq.2)}$$

where fermionic and bosonic parts $H_F, H_B$ are Hubbard like and interaction term has a form resembling electron-phonon interaction.

In the second order of perturbation theory an interaction term also becomes Hubbard-like [102]

$$H_{FB} = U_{FB} \sum n_{iF} n_{iB} \quad \text{(Eq.3)}$$

and depending upon the sign of $U_{FB}$ describes either Fermi-Bose mixture with repulsion [34] or with attraction [103] between fermions and bosons.

The Fermi-Bose mixture of spinons and holons interacting via a strong confinement potential [8,65] created by an AFM string [113,114] was considered in [115] for the underdoped high $T_C$ cuprates. The main idea was to describe the superconducting pairing in the strongly underdoped region of the t-J model on the language of the BCS-BEC crossover for the pairing in the d-wave channel of two spin polarons (two composite holes or two strings), where the critical temperature in the BCS phase is governed by (Eq. 1) and its spin-polaronic generalization [40,41]. The crossover however is probably not smooth, containing quantum critical point (QCP) or even intermediate phases in between BCS and BEC phase [115].

Note that non-phonon mechanisms of superconductivity can easily explain superconductivity with the critical temperatures $T_C$=100 K typical for cuprate perovskites. However to get $T_C$: (250–260) K typical for metallic hydrogen alloys another type of models such as Fermi-Bose mixture model or the Fano shape resonance mechanism with exchange interaction between a first BCS and a second condensate in the BCS-BEC crossover regime model is needed.

### 3. The 3D and 2D models of BCS-BEC crossover.

Note that to some extent the Femi-Bose mixture model naturally appears in the most difficult intermediate part of the phase diagram for the BCS-BEC crossover [116,21] between extended Cooper pairs (dilute BEC regime) and local pairs (or bipolarons [117,118]) which correspond to the weakly repulsive Bogoliubov gas [119] of composed bosons (see Fig. 4).

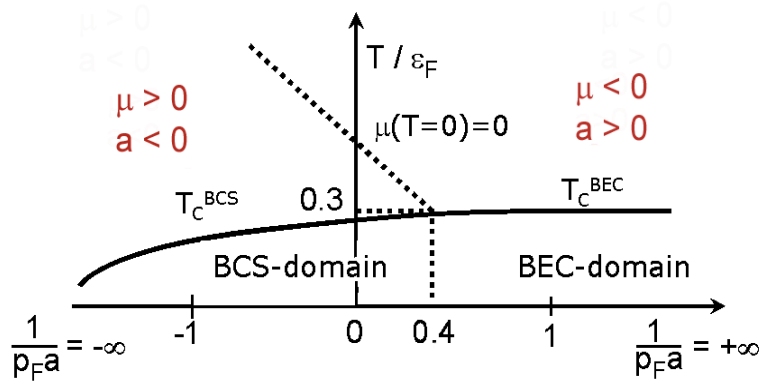

**Figure 4.** The typical phase-diagram of BCS-BEC crossover between extended Cooper pairs and local pairs in the 3D Fermi gas [116, 21].

Namely for the dilute Bose gas we have two distinct temperatures $T^*$ which corresponds to the formation of local pairs and $T_C$ of their Bose-condensation. For intermediate temperatures:

$$T_C \ll T \ll T^*, \qquad (\text{Eq. 4})$$

where the critical temperature of BEC

$$T_C = 3.31 \frac{(n/2)^{2/3}}{2m}[1 + 1.3 a_{2-2} n^{1/3}] \approx 0.2 E_F \qquad (\text{Eq. 5})$$

is mostly governed by Einstein formula [120] for the number of pairs density $n_B = n/2$ and the mass of the pair $m_B = 2m$. However there are nontrivial corrections to Einstein results connected with weakly repulsive interaction between the local pairs (between the dimers) [121]. This interaction is defined by the dimer-dimer scattering length $a_{2-2} = 0.6|a| > 0$ [122,123], where $a$ is an s-wave scattering length for particle-particle interaction. In the same time the crossover (Saha) temperature [124, 125, 21].

$$T^* \sim \frac{|E_b|}{3/2 \ln(\frac{|E_b|}{E_F})}, \qquad (\text{Eq. 6})$$

where $|E_b| = 1/ma^2$ is an absolute value of the binding energy of a local pair, $E_F = \frac{p_F^2}{2m}$ - is the Fermi energy, $p_F$ - is Fermi momentum.

Note that in the 2D case we have the similar estimate for the Saha temperature [125,126,127]

$$T^* \sim \frac{|E_b|}{\ln(\frac{|E_b|}{E_F})}, \qquad (\text{Eq. 7})$$

In the same time the BEC critical temperature is given by Fisher-Hohenberg theory [128] for weakly repulsive 2D Bose gas. According to [128]

$$T_C \sim \frac{E_F}{4\ln(1/f_{2-2})}, \qquad (\text{Eq.8})$$

where $f_{2-2} \sim 1/\ln(1.6|E_b|/E_F)$ [129] describes the repulsive interaction between the dimers in 2D.

For the intermediate temperatures in (Eq. 4) we have a new state of matter, namely the normal (non-superconducting) metal. According to [126, 127] it has very peculiar transport and thermodynamic properties with a resistivity behaving in semiconducting fashion $R \propto \sqrt{T}$.

In the same time in the intermediate coupling case (for the values of the gas parameter $1 \leq ap_F \leq 3$ [21]) the binding energy of the pair $|E_b|$ becomes comparable with the Fermi-energy $E_F$. When we increase the density we finally reach the limit $|E_b| \leq 2E_F$, and the local pairs start to touch each other. As a result some of the local pairs are crushed and (in the framework of e.g. 3D Fermi gas with attraction or attractive-U Hubbard model) we have a Fermi-Bose mixture of paired and unpaired electrons already at low temperatures.

With some precaution we can also describe the low temperature phase in case of intermediate coupling on the language of the Bose-Bose mixture of two coexisting Bose condensates one in the limit of extended pairs and a second one in the limit of almost local pairs.

## 4. Fano resonaces in multigap BCS-BEC superconductivity at Lifshitz transitions

Major schools of unconventional mechanisms for high temperature superconductivity have been based on the assumption of a "single electronic component" forming the superconducting condensate. The resonant valence bond theory proposed the condensation of preformed electronic pairs in a doped Mott insulator [3], the spin fluctuation theory proposed the pairing mediated by spin fluctuations in a strongly correlated metal [4]. The conventional BCS theory as well consider a single electronic component both in the weak coupling as well as in the strong coupling limit. In these standard theories the superconductor is formed by a single quantum condensate. In order to apply the BCS theory to a large variety of superconductors the complex band structure of metals with multiple Fermi surfaces is reduced to a single band theory using the "dirty limit approximation". However these "standard theories" have been falsified by the results of material science which has found high temperature superconductivity not in simple elemental metals which can be reduced to a single electronic component in a single Fermi surface but in complex multi-components materials with coexistence of multiple distinguishable electronic components with different orbital symmetries at the Fermi level with multiple different Fermi surfaces going from A15 intermetallics, doped cuprate perovskites, diborides, iron based superconductors, pressurized hydrides. In these systems different pairs in multiple portions of the Fermi surfaces condense in a single macroscopic coherent quantum state formed by multiple superconducting quantum condensates with different symmetries. While "exchange interaction" between quantum condensates is not considered and neglected in standard theories of superconductivity it becomes a key ingredient in multigap superconductors. The role of exchange interaction in the formation of a quantum heterogeneous matter made of different components was first faced in quantum mechanics in theories proposed to explain the formation of nuclear matter made by protons and neutrons.

The attractive exchange interaction between bosonic pairs of protons and bosonic pairs of neutrons in Quantum Mechanics was proposed in 1933 by Majorana [130,131] after Heisenberg proposed the well known repulsive exchange interaction [132-133] between fermionic particles. Majorana demonstrated that symmetry played a vital role in bosonic or fermionic systems. In fact his force and the one of Heisenberg made very different predictions because of their different symmetry properties. The Majorana exchange interaction opened the road map for both the interaction boson model [134] and the interaction boson fermion model [135], and today it is at the base of the research to extract simplicity in complexity applied to a large spectrum of systems where the focus of the research is to identify the dynamical complex symmetry [136] as shown in Figure 5 for the triangular hyperbolic tiling, and it was extended to the case of high temperature superconductors with multiple gaps [137] and in molecular systems. It is now clear that more complex is the system ,the more important is to search for its symmetry as for example in the *vibron model* described by the so-called(4)u algebra and in the *electron-vibron model* described by even more complicated superalgebra (4 / )u1 [136].

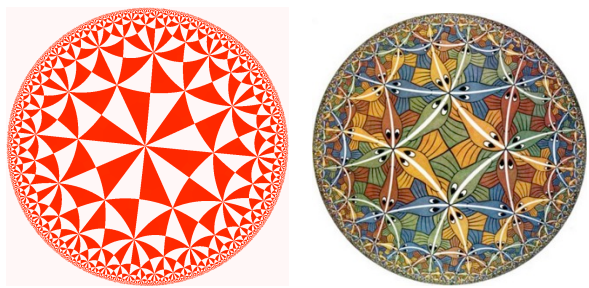

**Figure 5.** Tessellation of the hyperbolic Poincare' plane: showing the (6,4,2) triangular hyperbolic tiling (left side) that inspired the Escher *Circle limit III* pantin*g (right side) (From M.C. Escher, Circle Limit III, 1959).*

Fano introduced the configuration interaction between a closed channel and an open scattering channel in Quantum Mechanics in his famous paper published by Nuovo Cimento in 1935 [138]. He expanded and improved his theory in the 1961 Physical Review paper [139]. The idea of exchange interaction between pairs formed in the cloud of fermionic particles (which he called "pions") and a localized boson, an extension of Fano resonance, was developed in the quantum field theory of many body systems by Tomonaga [140]. The Fano resonances have been classified as a *"Feshbach resonance"* if scattering length of the system in the closed channel is *negative* and it gives a well localized boson pair at negative energy; or as a *"shape resonance"* if the scattering length of the system in the closed channel is positive and it is a *quasi-stationary state* at positive energy degenerate with the continuum [141].

The Bose fermion or Fermi-Bose model for high temperature superconductors is therefore described as the case of *Feshbach resonance* driven by Majorana exchange interaction between bosonic pairs making a BEC condensate with cooper pairs making a BCS condensate [142-144] which is described in the right panel of figure 6.

In 1993 an alternative theoretical scenario for high temperature superconductors driven by a Fano resonances of the type called shape resonances [77,80,145-150] was proposed. where the Majorana exchange term is in action between a cooper pairs in the BCS condensate and the second condensate of polaronic pairs in the intermediate coupling regime forming a condensate in the BCS-BEC crossover which is described in the central panel of figure 6. The theoretical proposal was based on compelling experimental evidence collected from the study of the structure of the 91K $Bi_2Sr_2CaCuO_{8+y}$ superconductor that bulk superconductivity emerges in a stack of 2D layers of a strongly correlated electron liquid confined in finite domains of incommensurate modulated aperiodic lattice where two electronic components coexist: a first Fermi liquid and a second polaronic liquid coexisting with a polaronic generalized Wigner charge density wave [145-147].

The "*shape resonance* driven by Majorana exchange interaction between a first condensate in the BCS-BEC crossover and cooper pairs in a BCS condensate appears by tuning the chemical potential to the electronic topological transition called Lifshitz transition in a correlated interacting low dimensional metal [148-150]. This non-standard and unconventional superconductivity scenario is completely beyond all BCS approximations and it was described theoretically by solving numerically the Bogoliubov anisotropic gaps equations at the same time as the density equation, called Bianconi-Perali-Valletta (BPV) approach [77, 80,148-150].

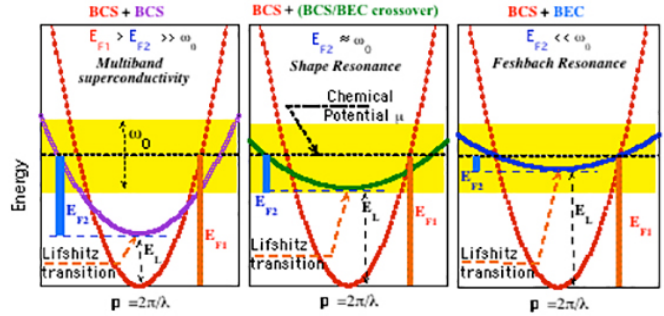

**Figure 6.** Three typical cases of non-standard superconducting theories of multigap superconductivity where Majorana exchange interaction between two different electronic components with different symmetry plays a key role: *left*, the case of two BCS condensates in two large Fermi surfaces; *center*: the case of Fano shape resonance between a first BCS condensate and a second condensate in the BCS/BEC crossover, *right:* the case of Feshbach resonance between a first BCS condensate and a second condensate in the BEC regime typical of the boson-fermion models.

The BPV theory predicts that the critical temperature is a strong function of the physical parameters like strain, doping, pressure, uniaxial pressure, charge density, gate voltage etc. which tune the system near the Lifshitz transition with a strongly anisotropic resonance of the superconducting critical temperature and the superconducting gap with Fano line-shape not predicted by standard single band theories.

The "shape resonance" superconducting phase in proximity of two Lifshitz transitions is supported in cuprates by the fact that the maximum critical temperature occurs at optimum doping δ=0.16 where the Fermi surface is made of disconnected Fermi arcs in the proximity of two topological electronic transitions. The first one is at δ=0.21 where for larger doping a large circular Fermi surface appears and the second for δ=0.125 where the Fermi surface shrinks to only four very short Fermi arcs in the (π,π) direction with a maximum isotope coefficient [150]. Complexity emerged in the beginning of the XXI century, as the second key universal feature of cuprate superconducting perovskites [151-153]. Three key features of high temperature superconductors were established: i) mesoscopic and nanoscale phase separation [151(see also review article [154])], ii) multiple electronic components in the normal metallic phase [152], iii) the key role of anisotropic lattice strain driving the complexity [153] and coexisting multiple superconducting gaps spatially distributed at nanoscale. The phase separation scenario with spin, charge, lattice nanoscale puddles with the proliferation of interface filamentary space was called a *Superstripes* scenario i.e., a supersolid with a specific symmetry [155] (see also Fig. 6 for illustration).

The BPV theory was presented at the first Stripes conference in 1996, before the discovery of superfluid condensation of ultracold bosonic gases in 1998 and the superfluid condensation of fermionic gasses in 2002 driven by the Feshbach resonance mechanism. At the Rome Stripes conference in 2000 the key role of anisotropic misfit-strain and the strain-doping-temperature phase diagram in the proximity of a Lifshitz transition have been presented. Few month later in Dec 2000 the superconducting properties of MgB$_2$, a binary metallic non-magnetic system known since fifty years, have been measured in Japan by Akimitsu. MgB$_2$ was a layered system not expected to be a BCS superconductor, showing the record of T$_c$=40K for a binary alloy. It was rapidly claimed that it was a clear case of superconductivity driven by a Fano shape resonance near a Lifshitz transition for the appearing of a new spot of a sigma Fermi surface with the key role of a large amplitude zero point motion at low temperature [156-157]. The BPV model predicted and provided the first published explanation of the high critical temperature in diborides where the misfit strain and doping moves the chemical potential in the $Al_{1-x}Mg_xB_2$ and $Sc_{1-x}Mg_xB_2$ below the Lifshitz transition for the appearing the sigma Fermi surface for x>0.5 [156]. It was the first case of a verified prediction of the BPV theory in a completely different system [148-150]. The second key feature, *nanoscale phase separation*, was found by doping $MgB_2$ with aluminum or scandium substitution for Mg [158] which shifts the chemical potential toward the Lifshitz transition.

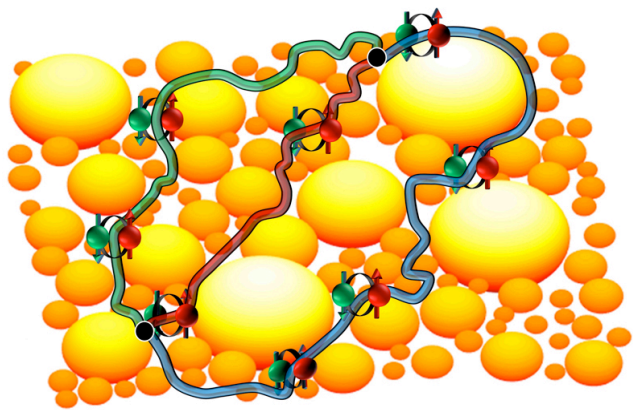

**Figure 7.** Superstripes landscape made of nanoscale striped puddles with a power law size distribution where the pathways of interface superconductivity connecting two points (where superconducting pairs shown in the figure) can be mapped to a hyperbolic space [81,155,175,176] providing a case of supersolid [39, 94] driven by strain and doping [152,153] near a critical point of a Lifshitz transition and of large zero point lattice fluctuations [151,155,157].

The second case of a different practical realization of the Fano shape resonance high temperature superconducting was reported in 2008 with the discovery of iron based superconductors [159-162] which are stacks of superconducting iron layers which show a strain doping phase diagram with maximum critical temperature in proximity of a Lifshitz transition [159-162].

After 2008 new microscopes using focused synchrotron radiation have shown an ubiquitous presence of nanoscale phase separation in high temperature superconductors in cuprates [81, 163-165] and in iron based superconductors [166-169] providing the theoretical prediction that nanoscale phase separation is an universal feature of systems showing Fano shape resonances in proximity of a Lifshitz transition [170-172] in the *superstripes* scenario [155] which is called by some authors with the generic name of intertwined or nematic phase. The theoretical efforts to understand the nanoscale phase separation in high temperature superconductors has pointed out the complexity of possible variable tilting of local structure in perovskites [171], the mixing of boson and fermions in the complex symmetry [172,173], the role of anisotropic pseudo-Jahn Teller vibronic interaction in the polaron formation [174] of these materials with the most recent results pointing out the role of a non-euclidean hyperbolic space (see Fig, 7) for the filamentary network at the interfaces of striped puddles in high temperature superconductivity [81,175-176].

**Conclusions**

In conclusion let us stress that recently discovered superconductivity in bilayer twisted graphene possibly also corresponds to the regime of BCS-BEC crossover between local and extended pairs in the d-wave channel (similarly to underdoped cuprates) but with anomalous chiral superconductivity of *d+id* type [177] which is in agreement with the phase diagram for idealized bilayer graphene in AB modification at low doping levels [20,22,23]. The discovery of superconductivity in graphene with many properties resembling the cuprates helps us to build the bridge between low temperature topological superconductors based on graphene [177] and bismuth [178-180] and vanadium oxides [181-184] which are very promising for superconducting nanoelectronics and quantum calculations (for creation of topologically protected qubits). The band structure calculations and recent experimental results show the evidence for multigap superconductivity in pressurized $H_3S$ hydrides as well as in the recently discovered high-temperature superconductor $LaH_{10}$ [84,85,86] where the superconducting dome occurs around the Lifshitz transition driven by pressure. Here the key role of Majorana exchange interaction driving the Fano shape resonance involve the new appearing small Fermi surface spots with strong pairing interactions where a strongly interacting set of states in the generalized strong-coupling approach predicted by the model of hydrogen bonds [185] is very important for hydrogen containing compounds. The evidence of shape resonance involving multiple set of electronic states in the strong coupling regimes has been stimulating a renewed interest to doped copper perovskites where the Hubbard models including multiple components showing a coexistence of coherent quasiparticles, incoherent states and stripe order with nondispersive quasiparticles close to the Fermi energy [186,187]. Finally the physics of XXI century is reaching a universal understanding of the emergence of quantum coherence of high temperature which could be manipulated by controlling the formation of hyperbolic quantum geometries in a nanoscale phase separation superstripes scenario (see Fig.7) with interacting networks of multiple components [81,155,175,176,188] by self organization of local structural tilts [171], the polaronic interaction [189], the zero point motion [74-76] and the mobile dopants [163,190], controlled by anisotropic strain [153,159,189] in quantum complex materials which can be observed by novel methods using a scanning sub-micron x-ray beam of focused synchrotron radiation [163].

In conclusion the maximum superconducting critical temperature has been seen to be reached at near a Lifhsitz transition with a quantum condensate in the new appearing small Fermi surface is in the BCS-BEC crossover predicted since 1996 [145-149] was first verified in $MgB_2$ in 2001 (191,192). The key role of Feshbach resonance driven by Majorana exchange interaction to stabilize the quantum condensates was later confirmed also by the realization of a quasi pure BEC condensate in fermionic ultracold gases driven by Feshbach resonance [193] the physical properties of a quantum condensate in the BCS-BEC crossover in a single band case has been discussed for the case of ultracold gases [194]. Following the evidence for the experimental realization of the transition of Feshbach resonances in ultracold fermionic gases from negative to positive scattering length [195] in the regime of superconducting Fano shape resonances as in magnesium diboride [156-158], the theoretical interest focused in the BCS-BEC crossover first for a single electronic component [196] and recently for two electronic components [197].

**Acknowledgements:** M. Yu. K. is grateful to RFBR grant **N** 17-02-00135 and thanks the Program of Basic Research of the National Research University "Higher School of Economics" for support.